\documentclass[epj]{svjour}
\usepackage{graphics}
\def\Li{LiCu$_2$O$_2$}
\def\LiV{LiVCuO$_4$}
\def\beq{\begin{equation}}
\def\eeq{\end{equation}}

\begin{document}
\title{Microscopic mechanisms of spin-dependent electric polarization in 3d oxides}
\author{A.S. Moskvin\inst{1} \and S.-L. Drechsler\inst{2}
}                     
%
%
\institute{Ural State University, 620083 Ekaterinburg,  Russia \and Leibniz Institut f\"ur Festk\"orper- und Werkstoffforschung
Dresden, D-01171, Dresden, Germany}
\date{Received: date / Revised version: date}
%
\abstract{
We present a short critical overview of different microscopic models for nonrelativistic and relativistic magnetoelectric coupling including the so-called "spin current scenario", {\it ab-initio} calculations, and several recent microscopic approaches to a spin-dependent electric polarization in 3d oxides. 
} 
\maketitle
\section{Introduction}
\label{intro}
Since Astrov's discovery of the magnetoelectric (ME) effect in Cr$_2$O$_3$\cite{Astrov} several microscopic mechanisms of magnetoelectric coupling were proposed\cite{Fiebig}, however, the multiferroicity (see e.g., Refs.\cite{KimuraHur} and review articles  Refs.\cite{Fiebig,Khomskii}) has generated an impressive revival of the activity in the field. The microscopic origin of the magnetically driven electric polarization is the topic of an intense and controversial debate. 
Currently  two essentially different spin structures of net electric polarization in crystals are considered:  a bilinear {\it nonrelativistic symmetric} spin coupling  
\begin{equation}
	{\bf P}_s=\sum_{mn}{\bf \Pi}_{mn}^s({\bf S}_m\cdot {\bf S}_n)\,
	\label{TMS}
\end{equation}
and a bilinear {\it relativistic antisymmetric} spin coupling  
\begin{equation}
	{\bf P}_a=\sum_{mn}{\bf \stackrel{\leftrightarrow}{\Pi}}_{mn}^a \left[{\bf S}_m\times {\bf S}_n\right]\, ,
	\label{DM}
\end{equation}
respectively. If the first term  stems somehow or other from a spin isotropic Heisenberg exchange interaction, the second term does from antisymmetric Dzyaloshinsky-Moriya (DM) coupling.
These  effective spin-operator forms do not discriminate the "ionic" and  "electronic" contributions to magnetically driven ferroelectricity related with the off-center ionic displacements and the electron density redistribution, respectively. 
A microscopic quantum theory of the ME effect has not yet been fully developed, although several scenarios for particular materials have been proposed. Many authors consider that the giant multiferroicity requires the existence of sizeable atomic displacements and structural distortions driven by isotropic symmetric exchange coupling\cite{Chapon,Harris,delaCruz} or antisymmetric DM coupling\cite{Sergienko} ignoring, however, the fact that the  effects of nuclear displacements and electron polarization should be described on equal footing, e.g.,   in frames of the  well-known shell model of Dick and Overhauser \cite{shell} widely used in lattice dynamics.  Shell and core displacements may be of a comparable magnitude. The conventional shell model does not take into account the spin and orbital degrees of freedom,  hence it cannot describe the multiferroic effects. In fact, the displacements of both the atomic core and electron shell would depend on the spin surroundings producing an synergetic effect of spin-dependent electric polarization. Obviously, this effect manifest itself differently in neutron and x-ray diffraction experiments. Sorting out two contributions  is a key issue in the field.

The second harmonic generation (SHG) spectroscopy reveals a giant effect in compounds with magnetically driven ferroelectricity (TbMn$_2$O$_5$, MnWO$_4$) thus pointing to an electronic rather than ionic origin of the spontaneous  polarization\cite{Lottermoser}.
Interestingly, the exchange-induced electric dipole moment (\ref{TMS}), derived many years ago \cite{TMS}(see, also Ref.\cite{Druzhinin}), seems to be a natural electronic mechanism for giant multiferroicity. However,  the second, or exchange-relativistic electric dipole moment (\ref{DM}), despite its visible weakness, is at present frequently addressed to be the leading contributor to electronic multiferroicity, mainly due to the so-called "spin-current" mechanism\,\cite{Katsura1,Mostovoy}. To a large extent it is explained by two reasons. First, in a most part of multiferroics the electric polarization is observed when the magnetic ordering is of a type that breaks chiral symmetry, e.g., spiral or helical order with a nonzero "spin-current" $\propto \left[{\bf S}_m\times {\bf S}_n\right]$. Second, such a mechanism allows us to easily predict  the direction of the ferroelectric polarization for certain helical spin structures: ${\bf P}_a\propto \sum_{mn}\left[{\bf R}_{mn}\times \left[{\bf S}_m\times {\bf S}_n\right]\right]$ ($(\Pi_{mn})_{ij}\propto\epsilon_{ijk}({\bf R}_{mn})_k$)\,\cite{Katsura1}, or ${\bf P}_a\propto \left[{\bf e}\times {\bf Q}\right]$ (${\bf e}$ is the helix spin rotation axis, ${\bf Q}$ is the helix wave vector)\,\cite{Mostovoy}. 
It is worth noting that phenomenological theory by Mostovoy\,\cite{Mostovoy} implies a coexistence of two independent order parameters ${\bf P}({\bf r})$ and ${\bf M}({\bf r})$ coupled by the magnetoelectric energy: $\Phi _{me}({\bf P}({\bf r}),{\bf M}({\bf r}))=\gamma {\bf P}\left[{\bf M}({\bf \nabla}\cdot{\bf M})-({\bf M}\cdot{\bf \nabla}){\bf M}\right]$, while for the multiferroics under consideration we deal with ${\bf M}({\bf r})$ to be a leading, or first order parameter, and ${\bf P}({\bf r})$ as a second  order parameter which is expressed in terms of the former one. Furthermore, his phenomenological theory does not explain neither the origin nor the magnitude of the ME coupling parameter $\gamma $. 

"Ferroelectricity caused by spin-currents" has established itself as one of the leading
paradigms for both theoretical and experimental investigations in the field of
strong multiferroic coupling. However, a "rule" that chiral symmetry needs to be broken in order to induce a ferroelectric moment at a magnetic phase transition is questionable\,\cite{Betouras}.
There are notable exceptions, in particular, the manganites RMn$_2$O$_5$, HoMnO$_3$, where  a ferroelectric polarization can appear without any indication of a magnetic chiral symmetry breaking\,\cite{Chapon,Sergienko1}, and delafossite CuFe$_{1-x}$Al$_x$O$_2$, where the helimagnetic ordering generates a spontaneous electric polarization parallel to the helical axis\,\cite{CuFeAlO}, in sharp contrast with the prediction of the spin current model.  Moreover, there are increasing doubts whether weak exchange-relativistic coupling can generate the giant electric polarization observed in some multiferroics. 

Another point of hot debates around the microscopic origin of ME coupling is related with recent observations of a multiferroic behaviour concomitant the incommensurate  spin spiral ordering in  chain cuprates LiVCuO$_4$ \cite{Naito,Naito1,Schrettle} and LiCu$_2$O$_2$   \cite{Cheong,Seki,Rusydi}. At first sight, these  cuprates seem to  be  prototypical examples of 1D spiral-magnetic ferroelectrics revealing the $relativistic$ mechanism of  "ferroelectricity caused by spin-currents"\cite{Katsura1}. However, these, in particular, LiCu$_2$O$_2$ show up a behavior which is obviously counterintuitive within the framework of spiral-magnetic ferroelectricity\cite{Cheong}. Furthermore, quantum helimagnets  NaCu$_2$O$_2$ and Li$_2$ZrCuO$_4$ with a very similar CuO$_2$ spin chain arrangement do not reveal signatures of a multiferroic behavior. Thus, there is no clear understanding connecting all these striking properties.

Despite its popularity, the original "spin-current" model\,\cite{Katsura1} and its later versions \cite{Jia1,Jia2} (see also Ref.\,\cite{Nagaosa}) seem to be  questionable  as the authors proceed with an unrealistic scenario. Indeed, when addressing a generic centrosymmetric  M$_1$-O-M$_2$ system they groundlessly assume an effective spin polarization Zeeman field $\frac{U}{2}{\bf m}_i$ (${\bf m}_i $ is a local magnetic moment)  to align noncollinearly the spins of 3d electrons and to provide a nonzero value of the two-site spin current $\left[{\bf S}_1\times {\bf S}_2\right]$. The energy separation $U$ originates from the local Coulomb repulsion and the Hund coupling in the magnetically ordered phase\,\cite{Katsura1,Jia1,Jia2,Nagaosa}. Such an assumption goes beyond all the thinkable perturbation schemes and leads to an unphysically large effect of breaking of a spatial symmetry induced by a spin configuration that manifests itself in an emergence of a nonzero electric dipole moment for an isolated centrosymmetric M$_1$-O-M$_2$ system\,\cite{Katsura1}.

 Size of the macroscopic polarization {\bf P} in nonmagnetic ferroelectrics computed by modern {\it ab-initio} band structure methods agrees exceptionally well with the ones observed experimentally. However, the state of the art {\it ab-initio} computations for different multiferroics: manganites HoMnO$_3$\cite{Picozzi},  TbMn$_2$O$_5$\cite{Wang}, HoMn$_2$O$_5$\cite{Giovanetti}, spin spiral chain cuprates \LiV \, and \Li\,\cite{Xiang} yield data spread within one-two orders of magnitude with  ambiguous and unreasonable values of polarization depending on whether these make use of theoretical or experimental structural data or different values of the correlation parameters.  The basic starting points of the current versions of such spin-polarized approaches as the LSDA seem to exclude any possibility to obtain a reliable quantitative estimation of  the spin-dependent electric polarization in multiferroics\,\cite{Moskvin-mechanism}. We should  emphasize two weak points of so-called {\it first-principle calculations} which appear as usual to be well forgotten in the literature. First, these approaches imply the spin configuration  induces immediately the appropriate breaking of spatial symmetry that makes the symmetry-breaking effect of a spin configuration to be unphysically large. It is worth noting that the spin-current scenario\,\cite{Katsura1} starts with the same LSDA-like assumption of unphysically large symmetry-breaking spin-magnetic field.  
   Conventional schemes imply just the opposite, however, a physically more reasonable picture when the charge and orbital anisotropies induce a spin anisotropy. 
   Second, the {\it first-principle calculations} neglect quantum fluctuations, that restricts drastically their applicability to a correct description of the ME coupling derived from the high-order perturbation effects.

 Below we present a short overview of different microscopic approaches to spin-dependent electric polarization.  In Sec.II we address a systematic standard microscopic theory   which implies the derivation of effective spin operators for  nonrelativistic and relativistic contributions to electric polarization of the generic three-site two-hole cluster such as Cu$_1$-O-Cu$_2$. In Sec.III we address an alternative approach based on the parity breaking exchange coupling and the exchange-induced electric polarization effects. Sec.IV is focused on the microscopic origin of the multiferroic behaviour observed in the edge-shared CuO$_2$ chain compounds \LiV \, and \Li.
In Sec.V we draw attention to anomalous magnetoelectric properties of the electron-hole dimers to be precursors of the disproportionated phase which droplets can survive even in nominally pure undoped manganites.
 
\section{Spin-dependent electric polarization in a three-site M$_1$-O-M$_2$ cluster}

Generic three-site  M$_1$-O-M$_2$ cluster  forms a basic element of the crystalline and electron structure of 3d oxides. A realistic perturbation scheme needed to  describe properly  the active M 3d and O 2p electron states implies strong intra-atomic correlations, the comparable effect of crystal field, the quenching of orbital moments by a low-symmetry crystal field, account for the dp-transfer up to the fourth order effects, and a rather small spin-orbital coupling. To this end we make use of a technique suggested in refs.\,\cite{DM-JETP,DM-PRB}. 
 
 
\subsection{Three-site two-hole M$_1$-O-M$_2$ cluster}  
For illustration, below we address a  three-site (Cu$_1$-O-Cu$_2$) two-hole system typical for cuprates with a tetragonal Cu on-site symmetry and a Cu 3d$_{x^2-y^2}$ ground  states.    
 We start with the construction of the spin-singlet and spin-triplet wave functions for the  system taking account of the p-d hopping, on-site hole-hole repulsion, and crystal field effects for  excited configurations $\{n\}$ (011, 110, 020, 200, 002) with different hole occupation of Cu$_1$, O, and Cu$_2$ sites, respectively. The p-d hopping for Cu-O bond implies a conventional Hamiltonian
 \begin{equation}
 \hat H_{pd}=\sum_{\alpha \beta}t_{p\alpha d\beta}{\hat p}^{\dagger}_{\alpha}{\hat d}_{\beta}+h.c.\, ,
 \label{Hpd}
\end{equation}
 where ${\hat p}^{\dagger}_{\alpha }$ creates a hole in the $\alpha $ state on the oxygen site, while  ${\hat d}_{\beta}$
annihilates a hole  in the $\beta $ state on the copper site; $t_{p\alpha d\beta}$ is a pd-transfer integral. 
 For basic 101 configuration with two $d_{x^2-y^2}$ holes localized on its parent sites we arrive at a perturbed wave function as follows
 \begin{equation}
\Psi_{101;SM}=\eta_{S}[\Phi_{101;SM}+\sum_{\Gamma\{n\}\not=101}c_{\{n\}}(S\Gamma)\Phi_{\{n\};\Gamma SM}],
	\label{Psi}
\end{equation}
where the summation runs both on different configurations and different orbital $\Gamma$ states; 
\beq
\eta_{S}=[1+\sum_{\{n\}\Gamma}|c_{\{n\}}(S\Gamma)|^2]^{-1/2}
\label{norm}
\eeq
is a normalization factor. It is worth noting that the probability amplitudes, or hybridization parameters, $c_{\{011\}}$, $c_{\{110\}}\propto t_{pd}$, $c_{\{200\}}$, $c_{\{020\}}$, $c_{\{002\}}\propto t_{pd}^2$. 
To account for relativistic effects in the three-site cluster one should incorporate the spin-orbital coupling $V_{so}=\sum_i \xi_{nl}({\bf l}_i\cdot{\bf s}_i)$ both for 3d- and 2p-holes with a single particle constant $\xi_{nl}>0$ for electrons and $\xi_{nl}<0$ for holes.

In terms of the hole spins the conventional bilinear spin Hamiltonian for the Cu$_1$-O-Cu$_2$ system reads  as follows:
\begin{equation}
\hat H_s(12)=J(\hat{\bf s}_1\cdot \hat{\bf s}_2)+{\bf D}\cdot [\hat{\bf s}_1\times \hat{\bf s}_2]+\hat{\bf s}_1{\bf \stackrel{\leftrightarrow}{K}} \,\hat{\bf s}_2 \, ,\label{1}
\end{equation} 
 where $J\propto t_{pd}^4$ is an exchange integral, ${\bf D}\propto t_{pd}^4\xi_{nl}$ is a Dzyaloshinsky vector, ${\bf \stackrel{\leftrightarrow}{K}}\propto t_{pd}^4\xi_{nl}^2$ is a symmetric second-rank tensor of the anisotropy constants\,\cite{DM-JETP,DM-PRB}.    

\subsection{Nonrelativistic mechanism of spin-dependent electric polarization:local and nonlocal terms}
Projecting the electric dipole moment ${\bf P}=|e|({\bf r}_1+{\bf r}_2)$ on the spin singlet or triplet ground state of the two-hole system we arrive at an effective electric polarization of the three-center system  $\langle {\bf P}\rangle _{S}=\langle \Psi _{101;SM}|{\bf P}|\Psi _{101;SM}\rangle$ to consist of $local$ and $nonlocal$ terms, which accomodate the diagonal  and off-diagonal on the ionic configurations matrix elements, respectively\,\cite{Moskvin-mechanism}. The local contribution describes the redistribution of the local on-site charge density and can be written as follows:
$$
\langle {\bf P}\rangle ^{local}_{S}= |e||\eta_S|^2\big[({\bf R}_1+{\bf R}_2+({\bf R}_1+{\bf R}_{O})\sum_{\Gamma}|c_{110}(S\Gamma)|^2 
$$
$$
+({\bf R}_{O}+{\bf R}_2\sum_{\Gamma}|c_{011}(S\Gamma)|^2+2{\bf R}_{O}\sum_{\Gamma}|c_{020}(S\Gamma)|^2 
$$
\beq
+2{\bf R}_1\sum_{\Gamma}|c_{200}(S\Gamma)|^2+2{\bf R}_2\sum_{\Gamma}|c_{002}(S\Gamma)|^2 \big]-{\bf P}_0 \, ,
\label{local}
\eeq
where ${\bf P}_0=|e|({\bf R}_1+{\bf R}_2)$ is a bare purely ionic two-hole dipole moment.
Obviously, the net local  electric polarization  depends only on ${\bf R}_{ij}$ vectors (${\bf R}_{10},{\bf R}_{20},{\bf R}_{12}$). It is worth noting that the net local  electric polarization lies in the Cu$_1$-O-Cu$_2$ plane.
The nonlocal, or overlap contribution is related with the off-diagonal two-site matrix elements of ${\bf P}$\,\cite{Moskvin-mechanism}.

The effective electric polarization differs for the singlet and triplet pairing due to a respective singlet-triplet difference in the hybridization amplitudes  $c_{\{n\}}(S\Gamma)$. Hence we may introduce an effective nonrelativistic {\it exchange-dipole} spin operator 
\beq
{\bf \hat P}_s={\bf \Pi}({\bf \hat s}_1\cdot {\bf \hat s}_2)
\label{non}
\eeq
 with an {\it exchange-dipole} moment
\beq
{\bf \Pi}=\langle {\bf P}\rangle _{t}-\langle {\bf P}\rangle _{s}\, ,
\label{ts}
\eeq
which can be easily deduced from Exp. (\ref{local}).
The effective nonrelativistic exchange-dipole moment is determined by  competitive local and nonlocal contributions of several configurations\,\cite{Moskvin-mechanism}. 
It is worth noting that for the collinear Cu$_1$-O-Cu$_2$ bonding both contributions vanish. As a whole, the exchange-dipole moment vanishes, if the M$_1$-O-M$_2$ cluster has a center of symmetry.

It is worth  noting that we addressed only the charge density redistribution effects for Cu 3d and O 2p states and neglected a direct electronic polarization effects for the both metal and anion ions. These effects be incorporated to the theory, if other orbitals, e.g. $ns$- for oxygen ion, will be included in the initial orbital basis set. To proceed with these effects an alternative approach may be applied by using a generalized shell model\cite{Panov}.     
 
\subsection{Relativistic mechanism of the spin-dependent electric polarization}
 At variance with  a scenario by Katsura {\it et al.} \cite{Katsura1} we have applied a conventional procedure to derive an effective {\it spin operator}   for a relativistic contribution to the electric dipole moment in the three-site M$_1$-O-M$_2$ system like  a technique suggested in references \cite{DM-JETP,DM-PRB} to derive 
expressions for the Cu and O spin-orbital contributions to 
the DM coupling in cuprates. 

The spin-orbital coupling $V_{SO}$  for copper and oxygen ions drives the singlet-triplet mixing which gives rise to a relativistic contribution to electric polarization deduced from an effective spin operator, or an {\it exchange-relativistic-dipole} moment
 \beq
{\bf \hat P}=\frac{1}{2}{\bf \stackrel{\leftrightarrow}{\Pi}}{\bf \hat T}={\bf \stackrel{\leftrightarrow}{\Pi}}[\hat{\bf s}_1\times \hat{\bf s}_2]\, ,
\label{Pa}
\eeq
where 
\beq
\Pi _{ij}=-i\langle \Psi_s|P_i|\Psi_{tj}\rangle = \left(\langle \Phi_s|P_i|\Phi_{s}\rangle -\langle \Phi_t|P_i|\Phi_{t}\rangle\right)\frac{D_j}{J}
\eeq
 is an {\it exchange-relativistic-dipole} tensor ($\Psi_s$ and $\Psi_{tj}$ are spin singlet and spin triplet wave functions (\ref{Psi}), respectively). It is easy to see that this quantity has a clear physical meaning to be in fact a dipole matrix element for a singlet-triplet electro-dipole transition in our three-site cluster.
Taking into account equation (\ref{ts}), we arrive at a simple form for the exchange-relativistic-dipole moment as follows
\beq
{\bf \hat P}=-\frac{1}{J}{\bf \Pi}\left({\bf D}\cdot[\hat{\bf s}_1\times \hat{\bf s}_2]\right)\, .
\label{rel}
\eeq
It is worth noting that this vector lies in the Cu$_1$-O-Cu$_2$ plane and its direction, at variance with the spin-current model\,\cite{Katsura1},  does not depend on the spin configuration. Furthermore, we see that the both nonrelativistic and relativistic contributions to effective dipole moment in the Cu$_1$-O-Cu$_2$ system have the same direction: ${\bf P}_{s,a}\propto {\bf \Pi}^s_{12}$. In other words, in the both cases the spin-correlation factors, $(\hat{\bf s}_1\cdot\hat{\bf s}_2)$ and $[\hat{\bf s}_1\times \hat{\bf s}_2]$, do modulate a pre-existing dipole moment.
 The DM type exchange-relativistic-dipole moment (\ref{rel}) is believed to be a dominant relativistic contribution to the electric polarization in a Cu$_1$-O-Cu$_2$ cluster.
It is worth noting that the exchange-dipole moment operator (\ref{non}) and exchange-relativistic-dipole moment operator (\ref{rel}) are obvious counterparts of the Heisenberg symmetric exchange and DM antisymmetric exchange, respectively. Hence, the Moriya like relation\,\cite{Moriya}  $|\Pi_{ij}|\sim \Delta g/g |{\bf \Pi}|$  seems to be a reasonable estimation for the resultant relativistic contribution to electric polarization in  M$_1$-O-M$_2$ clusters. At present, it is a difficult and, probably, hopeless task to propose a more reliable and so physically clear estimate. 
 Taking a typical value of $\Delta g/g \sim 0.1$ we  estimate the maximal value of $|\Pi_{ij}|$ as $10^{-3}|e|\AA$($\sim 10^2 \mu C/m^2$) that points to the exchange-relativistic mechanism to be a weak contributor to a giant multiferroicity with ferroelectric polarization of the order of $10^3 \mu C/m^2$ as in TbMnO$_3$\,\cite{KimuraHur}, though it may be a noticeable contributor in, e.g., Ni$_3$V$_2$O$_8$\cite{Lawes}.

\section{Parity breaking exchange coupling and exchange-induced electric polarization} 
Along with many advantages of the three-site cluster model it has a clear imperfection  not uncovering a  direct role played by exchange coupling as a driving force to induce a spin-dependent electric polarization. Below we'll address an alternative approach starting with a spin center such as a MeO$_n$ cluster in 3d oxides exchange-coupled with a magnetic surroundings. Then the magnetoelectric coupling can be related to the spin-dependent electric fields generated by the spin  surroundings in a magnetic crystal. In this connection we should point out some properties of exchange interaction that usually are missed in conventional treatment of Heisenberg exchange coupling. Following the paper by Tanabe {\it et al.}\cite{TMS} (see, also Ref.\cite{Druzhinin}) we do start with a simple introduction to  exchange-induced electric polarization effects.

Let address the one-particle (electron/hole) center in a crystallographically centrosymmetric position of a magnetic crystal. Then all the particle states can be of a certain spatial parity, even (g) or odd (u), respectively. Having in mind the 3d centers we'll assume an even-parity ground state $|g\rangle$. For simplicity we restrict ourselves to only one excited odd-parity state  $|u\rangle$. The exchange coupling with the surrounding spins can be written as follows:
\begin{equation}
	{\hat V}_{ex}=\sum_n{\hat I}({\bf R}_n)({\bf s}\cdot {\bf S}_n),
\end{equation}
where ${\hat I}({\bf R}_n)$ is an orbital operator with a matrix
\begin{equation}
{\hat I}({\bf R}_n)=\pmatrix{I_{gg}({\bf R}_n)&I_{gu}({\bf R}_n)\cr
I_{ug}({\bf R}_n)&I_{uu}({\bf R}_n)\cr}.
\end{equation}
The parity-breaking off-diagonal part of exchange coupling can lift the center of symmetry and mix $|g\rangle$ and $|u\rangle$ states thus resulting in a nonzero electric dipole polarization of the ground state
\begin{equation}
	{\bf P}=\sum_{n}{\bf \Pi}_n({\bf s}\cdot {\bf S}_n)\, ,
\end{equation}
where 
\begin{equation}
{\bf \Pi}_n =	2I_{gu}({\bf R}_n)\frac{\langle g|e{\bf r}|u\rangle}{\Delta_{ug}} 
\end{equation}
with $\Delta _{ug}=\epsilon_u-\epsilon_g$. 
 It is easy to see that in frames of a mean-field approximation the nonzero dipole moment shows up only for spin-noncentrosymmetric surroundings, that is if the condition $\langle{\bf S}({\bf R}_n)\rangle=\langle{\bf S}(-{\bf R}_n)\rangle$ is broken. For an isotropic bilinear exchange coupling this implies a spin frustration.

It should be noted that at variance with the spin-current model\,\cite{Katsura1}   the direction of the exchange-induced dipole moment for $i,j$ pair does not depend on the direction of spins ${\bf S}_i$ and ${\bf S}_j$. In other words, the spin-correlation factor $({\bf S}_i\cdot {\bf S}_j)$ modulates a pre-existing dipole moment ${\bf \Pi}$ which direction and value depend on the Me$_i$-O-Me$_j$ bond geometry and orbitals involved in the exchange coupling. 
 
 The magnitude of the off-diagonal exchange integrals can sufficiently exceed that of a conventional diagonal exchange integral. Given reasonable estimations for the off-diagonal exchange integrals $I_{ug}\approx 0.1$ eV,  the $u-g$ energy separation $\Delta _{ug}\approx 2$ eV, the dipole matrix element  $|\langle g|e{\bf r}|u\rangle |\approx 0.1 \AA$, spin function $|\langle ({\bf s}\cdot {\bf S}_n)\rangle|\approx 1$ we arrive at an estimation of the maximal value of the electric polarization: $P\approx 10^4\,\mu C/m^{2}$. This estimate points to the exchange-induced electric polarization to be potentially the most significant source of magnetoelectric coupling for new giant multiferroics. 
 It is worth noting that the exchange-induced polarization effect we consider is particularly strong for the 3d clusters such as MeO$_n$ with an intensive low-lying electro-dipole allowed transition $|g\rangle \rightarrow |u\rangle $  which  initial and final states  are coupled due to a strong exchange interaction with a spin surroundings\,\cite{Moskvin-Pisarev}. This simple rule may be practically used to search for new multiferroic materials.
 
 The parity-breaking exchange coupling can produce a strong electric polarization of oxygen ions in 3d oxides which can be written as follows
\beq
	{\bf P}_O=\sum_{n}{\bf \Pi}_n(\langle{\bf S}_O\rangle\cdot {\bf S}_n)\, ,
\eeq
where ${\bf S}_n$ are the spins of the surrounding 3d ions, $\langle{\bf S}_O\rangle \propto \sum_{n}{\bf \stackrel{\leftrightarrow}{I}}_n{\bf S}_n$ is a spin polarization of the oxygen ion due to the surrounding 3d ions with ${\bf \stackrel{\leftrightarrow}{I}}_n$ being an exchange coupling tensor. It seems the oxygen exchange-induced electric polarization of purely electron origin has been too little appreciated in the current pictures of  multiferroicity in 3d oxides.

\section{Origin of multiferroic properties in the edge-shared C\lowercase{u}O$_2$ chain compounds}
\subsection{Cancellation of the spin-dependent electric polarization in perfect edge-shared C\lowercase{u}O$_2$ chains}

\begin{figure}
\resizebox{0.45\textwidth}{!}{%
\includegraphics{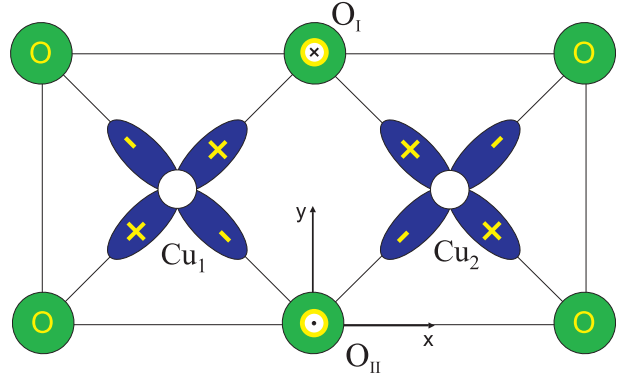}}
\caption{(Color online) The fragment of a typical edge-shared CuO$_2$ chain. Note the antiparallel orientation of the Dzyaloshinsky vectors in the Cu$_1$-O$_{I}$-Cu$_2$ and Cu$_1$-O$_{II}$-Cu$_2$ bonds both directed perpendicular to the chain plane\,\cite{DM-JETP}.}
\label{fig:1}      
\end{figure}

Recent observations of a multiferroic behaviour concomitant the incommensurate  spin spiral ordering in  cuprates LiVCuO$_4$ \cite{Naito,Naito1,Schrettle} and LiCu$_2$O$_2$   \cite{Cheong} with nearly perfect  edge-shared CuO$_2$ chains (see Fig.\,\ref{fig:1}) challenge the multiferroic community. From the viewpoint of the spin-current model, these  cuprates seem to  be  prototypical examples of the 1D spiral-magnetic ferroelectrics revealing the $relativistic$ mechanism of  "ferroelectricity caused by spin-currents"\,\cite{Katsura1}. 
However, as we see from  discussion above, isolated  perfect   edge-shared centrosymmetric CuO$_2$ chains  cannot produce a spin-dependent electric polarization both of nonrelativistic and relativistic origin.  Indeed, the net $nonrelativistic$ polarization of a spin chain  formed by Cu ions positioned at the center of symmetry  can be written as follows\,\cite{TMS}
\begin{equation}
{\bf P}_{eff}={\bf \Pi}\sum_{j=even}[({\bf S}_j\cdot {\bf S}_{j+1})-({\bf S}_j\cdot {\bf S}_{j-1})]	\, ,
\end{equation}
hence for a simple plane spiral ordering in perfect  edge-shared CuO$_2$ chains we arrive at a twofold cancellation effect  due to the zeroth value both of the ${\bf \Pi}$ and the spin-correllation factor in brackets.
A twofold cancellation effect takes place for the relativistic contribution (\ref{rel}) to the spin-dependent electric polarization as well, because both the exchange-dipole moment ${\bf \Pi}$ and  Dzyaloshinsky vector ${\bf D}$ turn into zero. Indeed, a specific symmetry of Cu$_1$-O-Cu$_2$ bonds in edge-shared CuO$_2$ chains (see Fig. \ref{fig:1}) results in a full cancellation of a net Dzyaloshinsky vector, though the partial Cu$_1$-O$_{I,II}$-Cu$_2$ contributions survive being of opposite signs\cite{DM-JETP,DM-PRB}.

The absence of a spin-dependent ferroelectric polarization in perfect  edge-shared CuO$_2$ chains is a simple corollary of its centrosymmetry.   The nonzero effect predicted by the spin-current model\,\cite{Katsura1} is related to an unphysical symmetry breaking engendered by a strong fictive "external" nonuniform field needed to align spirally the chain spins.
 Thus we may state that the edge-shared CuO$_4$ plaquettes arrangement in the CuO$_2$ chains  appears to be robust regarding the inducing of the spin-dependent electric polarization both of the nonrelativistic and relativistic origin. It means that we should look for the origin of puzzling multiferroicity observed in \LiV \, and \Li \, somewhere within the out-of-chain stuff.

\subsection{Nonstoichiometry and multiferroic behaviour of  edge-sharing C\lowercase{u}O$_2$ chain compounds LiVCuO$_4$ and LiCu$_2$O$_2$}

\begin{figure}
\resizebox{0.45\textwidth}{!}{%
\includegraphics{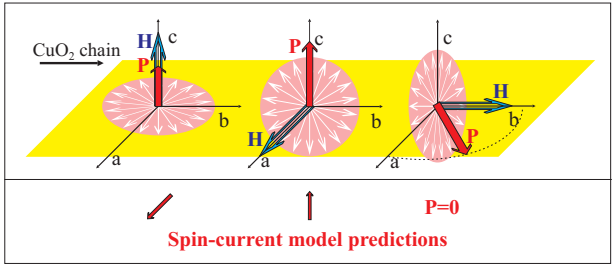}}
\caption{(Color online) Direction of ferroelectric polarization in \Li \, for different spin spiral plane orientation as observed by Park {\it et al.}\cite{Cheong} and predicted by the "nonstoichiometry" mechanism\,\cite{M+P+D}. For comparison the predictions of the spin-current model\,\cite{Katsura1,Mostovoy} are shown. }
\label{fig:2}      
\end{figure}

According to the spin-current theory\,\cite{Katsura1} a net electric polarization induced by a spin-spiral ordering in CuO$_2$ chains of quantum helimagnets \LiV \,,   NaCu$_2$O$_2$, \Li \,,  and Li$_2$ZrCuO$_4$ with a very similar CuO$_2$ spin chain arrangement is directed as shown in Fig.\,\ref{fig:1} with a magnitude proportional to $\sin\phi$, where $\phi$ is the pitch angle.  
Thus we should anticipate comparable values of a net chain electric polarization in these cuprates with pitch angles 85$^{\circ}$,  82$^{\circ}$, 62$^{\circ}$, 33$^{\circ}$, respectively\,\cite{Drechsler1}. However, these cuprates show up a behavior which cannot be explained within the framework of  spiral-magnetic ferroelectricity\,\cite{Katsura1,Mostovoy}. 
First, in accordance with a cancellation rule discussed above the quantum helimagnets  NaCu$_2$O$_2$ and Li$_2$ZrCuO$_4$  do not reveal any signatures of a multiferroic behavior while the both LiVCuO$_4$ and LiCu$_2$O$_2$ systems reveal a mysterious behavior with conflicting results obtained by different groups. Indeed, Yasui {\it et al.}\cite{Naito1}  claim that LiVCuO$_4$ reveals  clear deviations from the predictions of spin-current models\,\cite{Mostovoy,Katsura1} while Schrettle {\it et al.}\cite{Schrettle}  assure of its applicability. In contrast to LiVCuO$_4$,  LiCu$_2$O$_2$ shows up a behavior which is obviously counterintuitive within the framework of spiral-magnetic ferroelectricity\,\cite{Cheong}(see Fig.\,\ref{fig:2}). It is worth noting that at variance with Park {\it et al.}\cite{Cheong}, Naito {\it et al.}\cite{Naito} have not found any evidence for  ferroelectric anomalies in LiCu$_2$O$_2$.
 The ferroelectric anomaly in LiVCuO$_4$ reveals a magnitude ($P_a\approx 30\mu C/m^2$) comparable to that of the multiferroic Ni$_3$V$_2$O$_8$\,\cite{Lawes} while \Li \, shows up an order of magnitude lesser effect\,\cite{Cheong}. 
Recently we have shown that the  unconventional multiferroic behaviour observed in samples  of \LiV \, and \Li \, can have nothing to do with "spin-currents", this can be related with a nonstoichiometry in samples under consideration\,\cite{M+D,M+P+D}.  Their "multiferroicity"  can be consistently explained if one takes into account the $nonrelativistic$ exchange-induced electric polarization on the Cu$^{2+}$ centers substituting for the positions native for Li-ions in LiVCuO$_4$ and Cu$^{1+}$-ions in \Li, respectively\,\cite{M+D,M+P+D}. Such a mechanism does explain even subtle features of a multiferroic behavior observed in \LiV \, and \Li.  
These results raise a number of questions of great importance for  physics of magnetism and multiferroicity in the spin s=1/2 quantum matter of LiVCuO$_4$ and LiCu$_2$O$_2$.  
Whether a multiferroic behavior would be observed in stoichiometric samples with a regular arrangement of Cu$^{2+}$ and Cu$^{1+}$ ions? Recently, in order to exclude the nonstoichiometry as a source of a multiferroic behavior,  the Nagoya University group has prepared single-crystal samples of \Li\, with a controlled stoichiometry, which had "neither the atomic deficiency nor the mixing of Cu and Li atoms". The authors have tried to detect an electric polarization, however, at variance with earlier findings\,\cite{Cheong,Seki,Rusydi}  have not found any systematic data\,\cite{Sato}. Hardly visible bumps  of the capacitance for these samples ($\Delta C/C\approx 0.001$ at ${\bf E}\parallel {\bf c}$) observed at critical temperatures T$_{N_1}$ and T$_{N_2}$\,\cite{Yasui-09} cannot validate the spin-current theory\,\cite{Katsura1,Mostovoy}. These results are believed to  support strongly the  cancellation rule at work in the edge-shared CuO$_2$ chains and  the nonstoichiometry  as a source of a multiferroic behavior observed earlier both in \LiV \,\cite{Naito1,Schrettle} and  \Li\,\cite{Cheong,Seki,Rusydi}.

\section{Electron-hole dimers, electronic phase separation, and dielectric anomalies in  undoped  parent manganites}

\begin{figure}
\resizebox{0.45\textwidth}{!}{%
\includegraphics{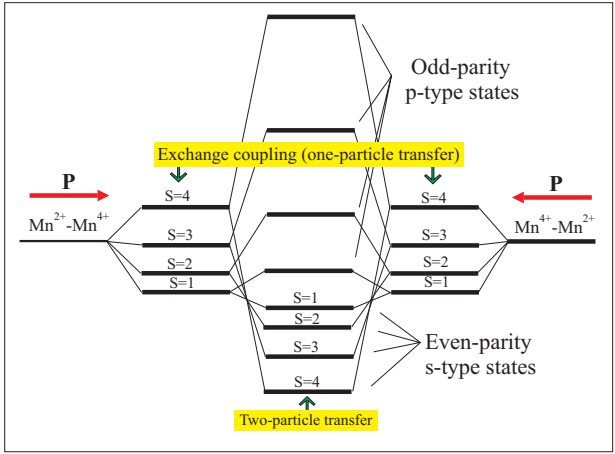}}
\caption{(Color online) Spin structure of the EH-dimer with a step-by-step  inclusion of one- and two-particle charge transfer. Arrows point to electric dipole moment for bare site-centred dimer configurations.}
\label{fig:3}      
\end{figure}

Even the nominally pure globally centrosymmetric  parent manganite LaMnO$_3$ exhibits a puzzling multiferroic-like behavior inconsistent with a simple picture of an A-type antiferromagnetic insulator (A-AFI) with a cooperative Jahn-Teller ordering.  Its anomalous properties are assigned to charge transfer instabilities and competition between insulating A-AFI phase and metallic-like dynamically disproportionated phase formally separated by a first-order phase transition at T$_{disp}$\,=\,T$_{JT}$$\approx$\,750\,K \cite{Moskvin-Mn-09}.  The unconventional high-temperature phase is addressed to be a specific electron-hole Bose liquid (EHBL) rather than a simple "chemically" disproportionated R(Mn$^{2+}$Mn$^{4+})$O$_3$ phase. New phase does nucleate as a result of the charge transfer (CT) instability and evolves from the self-trapped CT excitons, or specific EH-dimers, which seem to be a precursor of both insulating and metallic-like ferromagnetic phases observed in manganites.
The view of a self-trapped CT exciton to model a  Mn$^{2+}$-Mn$^{4+}$  pair is typical for a $chemical$ view of disproportionation, and is strongly oversimplified.  Actually we deal with an EH-dimer to be a dynamically charge fluctuating system of coupled electron MnO$_{6}^{10-}$ and hole MnO$_{4}^{8-}$ centers having been glued in a lattice due to strong electron-lattice polarization effects. In other words, we should proceed with a rather complex $physical$ view of disproportionation phenomena which first implies a  charge exchange reaction
\begin{equation}
\mbox{Mn}^{2+}+\mbox{Mn}^{4+} \leftrightarrow \mbox{Mn}^{4+}+\mbox{Mn}^{2+}\, ,	
\end{equation}
governed by a two-particle charge transfer integral
\begin{equation}
t_B=\langle \mbox{Mn}^{2+}\mbox{Mn}^{4+}|\hat H_B|\mbox{Mn}^{4+}\mbox{Mn}^{2+}\rangle \, ,	
\end{equation}
 where $\hat H_B$ is an effective two-particle (bosonic) transfer Hamiltonian, and we assume a parallel orientation of all the spins.
As a result of this quantum process the bare ionic states with site-centred charge order and the same bare energy $E_0$ transform into two EH-dimer states with an indefinite valence and bond-centred charge-order
\begin{equation}
|\pm \rangle =\frac{1}{\sqrt{2}}(|\mbox{Mn}^{2+}\mbox{Mn}^{4+}\rangle \pm |\mbox{Mn}^{4+}\mbox{Mn}^{2+}\rangle )	
\end{equation}
with the energies $E_{\pm}=E_0\pm t_B$. In other words, the exchange reaction restores the bare charge symmetry.
 In both $|\pm\rangle $ states the site manganese valence is indefinite with quantum  fluctuations between +2 and +4, however, with a mean value of +3.
Interestingly, in contrast with the  ionic states, the EH-dimer states $|\pm \rangle $  have both a distinct electron-hole and inversion symmetry, even parity ($s$-type symmetry) for $|+ \rangle $, and odd parity ($p$-type symmetry) for $|- \rangle $ states, respectively. Both states are coupled by a large electric-dipole matrix element:
\begin{equation}
\langle +|\hat {\bf d}|-\rangle =2eR_{MnMn}\, , 	
\end{equation}
where $R_{MnMn}$ is the Mn-Mn separation. 
 In a nonrelativistic approximation the spin structure of the EH-dimer will be determined by the isotropic Heisenberg exchange coupling  
$
V_{ex}=J\,({\bf S }_1\cdot {\bf S }_2),	
$
 and the two-particle charge transfer characterized by a respective transfer integral which depend on the spin states. 
 Both terms can be easily diagonalized in the net spin S representation so that for the energy we arrive at
\begin{equation}
E_S=\frac{J}{2}[S(S+1)-\frac{25}{2}]\pm \frac{1}{20}S(S+1)\,t_B\,,	
\end{equation}
where $\pm$ corresponds to two quantum superpositions $|\pm\rangle $ 
with $s$- and $p$-type symmetry, respectively.  It is worth  noting that the bosonic double exchange contribution formally corresponds to a ferromagnetic exchange coupling with $J_B=-\frac{1}{10}|t_B|$. 
We see that the cumulative effect of the Heisenberg exchange and the bosonic double exchange results in a stabilization of the S\,=\,4 high-spin
 (ferromagnetic) state of the EH-dimer provided $|t_B|>10J$ (see Fig.\ref{fig:3}) and  the S\,=\,1 low-spin (ferrimagnetic) state otherwise. The spin states with intermediate S values: S= 2, 3 correspond to a classical noncollinear ordering.
 
The EH-dimer reveals unconventional magnetoelectric properties. Indeed, the two-particle bosonic transport and respective kinetic contribution to stabilization of the ferromagnetic ordering  can be suppressed by a relatively small electric fields that makes the  EH-dimer to be a  promising magnetoelectric cell especially for the heavy rare-earth manganites RMnO$_3$ (R=Dy, Ho, Y, Er) with supposedly  a ferro-antiferro instability. In addition, a strong anisotropy of the dimer's electric polarizability  is noteworthy. In an external electric field the EH-dimers tend to align along the field.

Anomalous electric polarisability of the EH dimers and EH droplets that would result in dielectric anomalies in the EHBL phase and the phase-separated state of LaMnO$_3$. Indeed, such anomalies were reported recently both for poly- and single-crystalline samples of the parent LaMnO$_3$\cite{Mondal}. First of all, one should note the relatively high static dielectric constant in  LaMnO$_3$ at T\,=\,0 ($\varepsilon_0\sim 18-20$)  approaching to values typical for genuine multiferroic systems  ($\varepsilon_0 \approx 25$), whereas for the conventional nonpolar systems,  $\varepsilon_0$ varies within 1-5.
The entire $\varepsilon^{\prime}(\omega,T)$- T pattern across 77-900\,T has
two prominent features: (i)  near T$_N$ and  (ii)  near T$_{JT}$ to be essential signatures of puzzlingly unexpected multiferroicity, however, the intrinsic  electrical polarization  probably develops locally with no global ferroelectric order.
The observation of an intrinsic dielectric response  in the globally centrosymmetric LaMnO$_3$, where no ferroelectric order is possible due to the absence of off-centre distortion in MnO$_6$ octahedra cannot be explained in frames of a conventional uniform antiferromagnetic insulating  A-AFI scenario and agrees with the electronic A-AFI/EHBL phase separated state with a coexistence of the non-polar A-AFI phase and a highly polarizable EHBL phase\cite{Moskvin-Mn-09}.  

\section{Conclusion}

We have considered  several mechanisms of spin-dependent electric polarization in 3d oxides.
Starting with a generic  three-site two-hole cluster and a realistic perturbation scheme we have  deduced both nonrelativistic   and relativistic   contributions to the electric polarization. Nonrelativistic mechanism related to the redistribution of the local on-site charge density due to the $pd$ covalency and the exchange coupling is believed to govern the multiferroic behaviour in 3d oxides. The approach realized has much in common with the mechanism of the bond- and site-centered charge order competition (see, e.g. Ref.\cite{vdB}) though we started with the elementary $pd$ charge transfer rather than the $dd$ charge transfer. An alternative approach to the derivation of the spin-dependent electric polarization was considered which is based on the parity-breaking exchange coupling and the exchange induced  polarization. 
  As an actual application of the microscopic approach we discuss recent observations of multiferroic behaviour concomitant the incommensurate  spin spiral ordering in  s=1/2 chain cuprates \LiV \, and \Li. We argued that the multiferroicity observed in these nonstoichiometric cuprate samples has nothing to do with "spin currents" and can be consistently explained if one takes into account the nonrelativistic exchange-induced electric polarization on the Cu$^{2+}$ centers substituting for the positions native for the Cu$^+$-ions in \Li \, or the positions native for the Li$^+$-ions in \LiV, respectively. We argued that a charge transfer instability accompanied by nucleation of the electron-hole dimers and droplets in 3d oxides gives rise to a novel type of  magnetoelectric coupling due to a field-induced redistribution of  the electron-hole droplet volume fraction.

Financial support by the  RFBR through Grants Nos.  07-02-96047,  and  08-02-00633 (A.S.M.), and by the DFG under contract DR269/3-1 (S.-L.D)  are acknowledged. A.S.M. would like to thank Leibniz-Institut f\"{u}r Festk\"{o}rper- und Werkstoffforschung
Dresden, where part of this work was done, for hospitality.

\end{document}